\documentclass{article}
\usepackage[T1]{fontenc} 
\usepackage[utf8]{inputenc} 
\usepackage{ismir,amsmath,cite,url}
\usepackage{graphicx}
\usepackage{color}

\usepackage{booktabs} 
\usepackage{multirow}

\title{Long-Form Text-to-Music Generation with Adaptive Prompts: A Case Study in Tabletop Role-Playing Games Soundtracks}

\twoauthors
  {Felipe Marra} {Universidade Federal de Viçosa \\ Departamento de Informática \\ Viçosa, Minas Gerais, Brazil }
  {Lucas N. Ferreira} {Universidade Federal de Viçosa\\ Departamento de Informática \\ Viçosa, Minas Gerais, Brazil }

\def\authorname{F. Marra, L. Ferreira}

\newcommand{\bardo}{Babel Bardo}

\begin{document}

\maketitle
\begin{abstract}
This paper investigates the capabilities of text-to-audio music generation models in producing long-form music with prompts that change over time, focusing on soundtrack generation for Tabletop Role-Playing Games (TRPGs). We introduce \bardo, a system that uses Large Language Models (LLMs) to transform speech transcriptions into music descriptions for controlling a text-to-music model. Four versions of \bardo~were compared in two TRPG campaigns: a baseline using direct speech transcriptions, and three LLM-based versions with varying approaches to music description generation. Evaluations considered audio quality, story alignment, and transition smoothness. Results indicate that detailed music descriptions improve audio quality while maintaining consistency across consecutive descriptions enhances story alignment and transition smoothness. 


\end{abstract}
\section{Introduction}
\label{sec:introduction}

Recent text-to-audio music generation models such as MusicLM \cite{musiclm} and MusicGen \cite{musicgen} are capable of producing high-quality music in the audio domain that aligns with a given textual description. These models typically generate music autoregressively by predicting the next token from a context window, which limits the size of the signal they can model. While the context size is limited, these models can generate longer signals by sliding a context window through time. Regardless of this capability, they have mainly been evaluated with a fixed prompt and for relatively short music durations. For instance, MusicGen \cite{musicgen} was evaluated considering 30-second music pieces, each generated from a single music description. In this paper, we are interested in evaluating whether text-to-music models can maintain music quality while generating long music pieces, where music descriptions change over time.

It is important to evaluate text-to-music models considering long music pieces (greater than 30 seconds, for example) because many music production scenarios involve music durations longer than one can generate with a single short audio context window (e.g., pop music composition, jazz improvisation, soundtrack generation). One key problem of generating long sequences from a small context is that a model has to split the generation into multiple parts, ensuring that the independent parts are smoothly connected in the final composition. Moreover, one might change the initial prompt at any time step, steering the composition in a different direction, and the model must consider both the previous audio context and the new prompt.

\begin{figure}[t]
 \centerline{
 \includegraphics[width=1.\columnwidth]{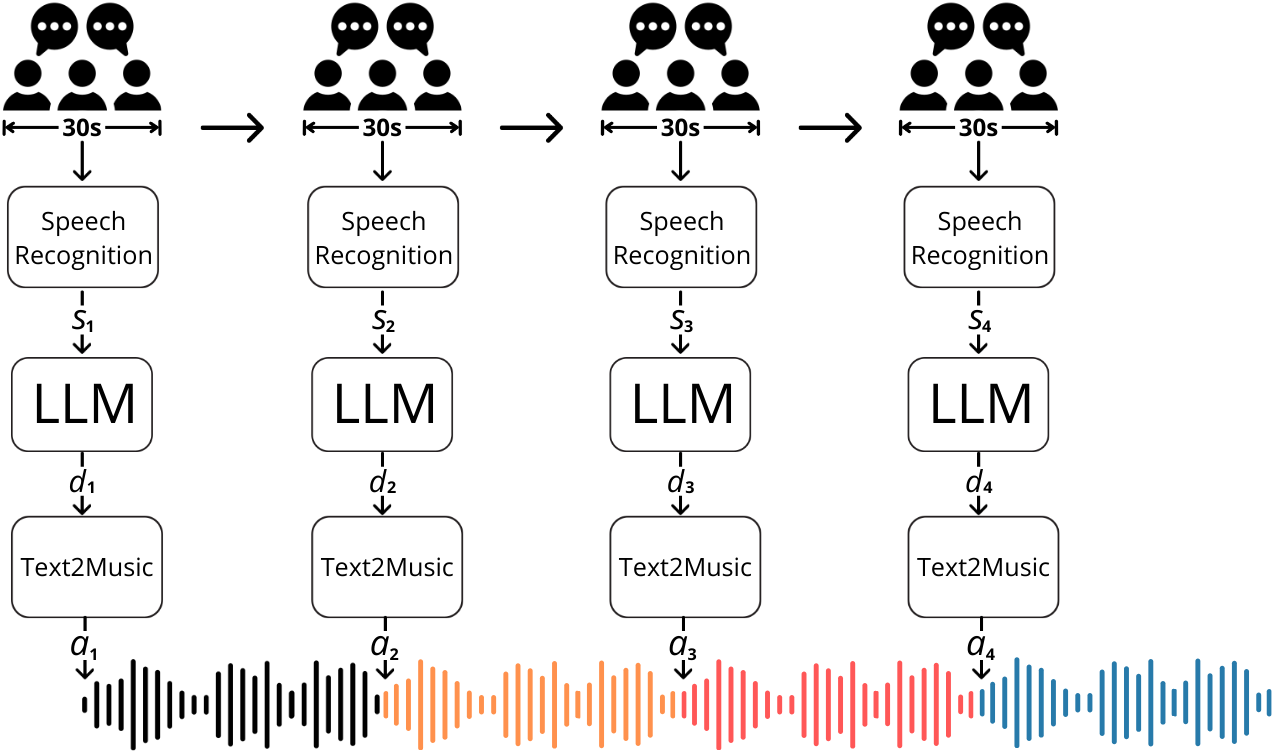}}
 \caption{At every 30 seconds of gameplay, \bardo~transcribes the players' speeches into a text $s_i$ using a Speech Recognition system and uses a Large Language Model (LLM) to map $s_i$ into a music description $d_i$ that matches the scene described by the players. This music description is given to a Text-to-Music system that generates a 30-second piece $a_i$ directly in the audio domain.}
 \label{fig:bardo_overview}
\end{figure}

In this paper, we investigate long generation with text-to-audio models in the context of Tabletop Role-Playing Games (TRPGs). In this scenario, a music generator takes speech as input and must generate music that matches the story being told by the players. We chose this problem because it inherently poses the challenge of long music generation, where prompts have to change over time to adjust for different story scenes. We also use TRPGs as a research object because TRPG players often enhance their gaming experience by manually selecting songs to play as background music \cite{bergstrom2014case}, which allows us to compare the results of a generator against a human baseline.

To investigate the capabilities of current text-to-music models in generating background music for TRPG stories, we've built a system called \bardo, which is inspired by Bardo Composer\cite{bardo_composer}, a system that generates symbolic music by transcribing players' speeches into text and conditioning an autoregressive model with the emotional tone of this text, as given by an emotion classifier. Different than Bardo Composer, \bardo~composes music directly in the audio domain by leveraging a Large Language Model (LLM) to transform the speech transcriptions into music descriptions every 30 seconds of gameplay. These descriptions are then given to a text-to-music model to generate a piece of music for that current moment of the story. Figure \ref{fig:bardo_overview} shows an overview of our system. \bardo~is also inspired by Hermann1\cite{hermmann1}, which uses LLMs and text-to-music models to generate soundtracks for films. 

We compared four different versions of \bardo~in two TRPG campaigns played on YouTube: Call of the Wild (in English) and \textit{O Segredo na Ilha} (in Brazilian Portuguese). The first version is our baseline and uses the speech transcriptions directly as prompts for a text-to-music model. All other versions use an LLM to transform the transcriptions into music descriptions. The second one follows the Bardo Composer approach and applies an LLM as an emotion classifier. The music description follows a template that is adjusted based on the emotion given by the LLM. The remaining two versions use the LLM to produce a complete music description; 
however, one generates a new description for every transcript, while the other can just continue the previously generated segment if the scene hasn't changed.

We evaluated our models according to audio quality, alignment with the story, and transition smoothness between transcriptions. Results suggest that while detailed music descriptions contribute to improved audio quality, maintaining consistency across consecutive descriptions helps achieve smoother transitions between musical segments. Furthermore, our findings indicate that emotion serves as an effective signal for aligning generated music with TRPG narratives.

\section{Related Works}
\label{sec:related_works}

This section reviews audio-based text-to-music models and previous soundtrack generation approaches, focusing on background music generation for TRPGs.

\subsection{Text-to-Music Models}

Text-to-music is the task of generating music pieces from music descriptions in textual format, e.g., ``70s punk rock song with fast tempo''. In recent years, various neural models have been proposed to solve this problem in the audio domain \cite{musiclm, musicgen, mousai}. For example, MusicGen \cite{musicgen} is an autoregressive transformer that operates on quantized audio units produced by the EnCodec \cite{encodec} audio tokenizer. It can be conditioned on textual descriptions using various text encoding methods (e.g., T5 \cite{t5}, FLAN-T5 \cite{flant5}, and CLAP \cite{clap}), or on melodic structures through an unsupervised approach utilizing chromagram information.

MusicLM \cite{musiclm} is another text-to-music model that extends AudioLM's \cite{audiolm} multi-stage autoregressive modeling approach by incorporating text conditioning, which is achieved by leveraging MuLan \cite{mulan} to project music and textual descriptions into a shared embedding space. Moûsai \cite{mousai} employs a two-stage cascading diffusion approach, where the first stage utilizes a novel diffusion magnitude-autoencoding (DMAE) technique to train a music encoder that compresses audio into a reduced representation. In the second stage, Moûsai implements text-conditioned latent diffusion (TCLD) to generate this reduced representation while conditioning on textual descriptions, enabling the model to produce music that corresponds to given text inputs. Other examples include commercial models such as Suno \cite{suno}, Mubert \cite{mubert}, and Riffusion \cite{riffusion}.

\subsection{Soundtrack Generation}

Soundtrack generation has been investigated in different mediums such as films \cite{hermmann1}, video games \cite{nesmvdb}, stories \cite{davis2014generating}, and others. This problem has been mainly studied in the symbolic domain. For example, Bardo Composer \cite{bardo_composer} generates background symbolic music for TRPGs by transcribing the players' speeches into text at every time step $t$ and feeding these transcriptions to a music classifier. The emotion given by this classifier is used to condition an autoregressive model that generates music by decoding a sequence using a variation of Stochastic Beam Search. \bardo~is similar to Bardo Composer because it also uses transcriptions of players' speeches to condition the music generation. However, it generates music in audio format by conditioning a text-to-music model with a music description at every time step $t$, instead of using a music emotion classifier to condition an autoregressive symbolic music model.

Another important related work is Herrmann-1 \cite{hermmann1}, which combines an LLM and a text-to-music model to generate background music for movie scenes. Herrmann-1 uses BLIP2~\cite{blip2} and CLIP~\cite{clip} to extract a textual description and the affective characteristics of the video, respectively. These characteristics are then provided as input to GPT-4~\cite{gpt4}, which generates a description of an appropriate music with the given characteristics. Finally, the description generated by GPT-4 is passed to MusicGen~\cite{musicgen}, which produces the background music in audio format. \bardo~is similar to Herrmann-1, because it also employs an LLM to generate music descriptions for a text-to-music, however, it takes text as input instead of videos. Moreover, in \bardo, the music descriptions change over time, whereas Herrmann-1 uses a single description for each video.

\section{Babel Bardo}

In TRPGs, players collaboratively construct a narrative through iterative cycles of scene descriptions, decision-making, and action resolutions. The game master presents scenarios and environmental details, to which players respond by declaring their characters' intended actions. These actions are then adjudicated using the game's rule system, often involving probabilistic elements, with the outcomes shaping the evolving storyline and informing subsequent player choices. This process creates an emergent and interactive storytelling experience.

\bardo~generates music for a TRPG story by iteratively transcribing the players' speeches into text and leveraging an LLM to produce a music description that is given to a text-to-music model, which in turn generates a piece of music. Formally, a story can be viewed as a sequence $S = \{s_1, s_2, ..., s_n\}$ of transcriptions, where each transcription $s_i$ is a string generated after $t$ seconds of gameplay. At each time step, \bardo~produces a music description $d_i$ by asking an LLM to generate one that aligns with the transcript $s_i$. The music description $d_i$ is then given as input to a text-to-music model that realizes the described music in audio format $a_i$. The text-to-music model also receives the previously generated audio $a_{i-1}$ as a conditional input, so it has to generate a music piece based on the description $d_i$ while continuing $a_{i-1}$. It is important to highlight that since we are using an LLM to produce music descriptions, \bardo~supports transcripts $s_i$ in any language.

To illustrate our proposed generative pipeline, consider the transcription $s_i$ = \textit{``You see a dragon in front of you. A battle will start!''}. \bardo~could produce a description $d_i$ = \textit{"A grand orchestral arrangement with thunderous percussion, epic brass fanfares, and soaring strings."}, which would be used by the text-to-music model together with the previously generated audio $a_{i-1}$ to produce the current audio segment $a_i$.

We evaluate four different approaches to combine an LLM and a text-to-music model to generate background music for TRPGs. All of them start with an initial prompt to condition the LLM with the generation task: \textit{``You are going to receive a series of Role-playing Game (RPG) video transcript excerpts from players' dialogues playing a campaign''}. After this initial setup, each model receives a sequence of transcripts $s_i$, and each approach employs the LLM in a different way to generate associated music descriptions $d_i = LLM(s_i)$.\vspace{0.5em}

\noindent
\textbf{\bardo~- Baseline (B)}. In this first version, \bardo~does not use the LLM to generate a music description. Instead, it uses the transcription $s_i$ directly as a prompt to the text-to-music model ($d_i = s_i$). \vspace{0.5em}

\noindent
\textbf{\bardo~- Emotion (E)}. This version behaves similarly to Bardo Composer and uses the LLM only as an emotion classifier. It processes the transcription $s_i$ with the following prompt: ``Classify each dialogue into one of the following emotions: Happy, Calm, Agitated, or Suspenseful.'' The LLM returns an emotion $e_i$, which is used to adjust the following pre-composed prompt $d_i$ = \textit{"Background music for a Role-playing Game (RPG) dialogue, with the following emotion: $e_i$"}. We used these four emotions because Bardo Composer \cite{bardo_composer} also used them. \vspace{0.5em}

\noindent
\textbf{\bardo~- Description (D)}. In this third version, \bardo~employs the LLM to generate a description $d_i = LLM(s_i)$ with the following prompt: ``For each transcript excerpt, describe a piece of background music that matches that excerpt.'' \vspace{0.5em}

\noindent
\textbf{\bardo~- Description Continuation (DC)}. This last version is similar to the previous one; however, it allows the LLM to keep the same description $d_i = d_{i-1}$ if the transcription $s_i$ is part of the same scene as $s_{i-1}$. This is achieved by producing a description $d_i = LLM(s_i)$ with the following prompt to the LLM: ``\textit{Determine whether this dialogue is from the same scene as the previous dialogue and based on this determination, either return the previous music description or generate a new one}''. This version is intended to help \bardo~keep a consistent soundtrack across a given scene by continuing $a_{i-1}$ without changing the description. \vspace{0.5em}

\section{Experiments and Results}

\begin{table}[t]
\centering
\setlength{\tabcolsep}{3pt}
\begin{tabular}{cccccc}

\toprule
\multicolumn{6}{c}{\textbf{$\downarrow$  FAD score}} \\
\toprule

\multirow{2}{*}{\textbf{TRPG}} & \multicolumn{4}{c}{\textbf{\bardo}} & \multirow{2}{*}{\textbf{Human}} \\
\cmidrule{2-5}

& \multicolumn{1}{c}{\textbf{B}}   & \multicolumn{1}{c}{\textbf{E}}   & \multicolumn{1}{c}{\textbf{D}}   & 
\multicolumn{1}{c}{\textbf{DC}}  & 
\\

\midrule

\multicolumn{1}{c}{\textbf{COTW}} & {9.66} & {5.99} & {6.25} & {\textbf{5.82}}  & {3.00}\\ 

\midrule

\multicolumn{1}{c}{\textbf{OSNI}} & {9.55} & {6.11} & {5.63} & {\textbf{5.13}} & {4.18}\\ 

\bottomrule

\end{tabular}
\caption{FAD for each \bardo~version in COTW and OSNI in contrast with Human music, i.e., the original soundtracks used by the players in both these TRPGs.}
\label{tab:fad_cotw}
\end{table}

We evaluate \bardo\footnote{\url{github.com/FelipeMarra/babel-bardo}} in the task of soundtrack generation for two different TRPG campaigns: Call of the Wild (COTW) and \textit{O Segredo da Ilha} (OSNI). The former is a Dungeons \& Dragons campaign played in American English and the latter is in Brazilian Portuguese. Both of them were played on YouTube. We used COTW because it was also used to evaluate Bardo Composer \cite{bardo_composer}. We've also included OSNI to evaluate \bardo's performance with a Latin American language. COTW is composed of 11 episodes, with a total of 6 hours and 37 minutes of gameplay—each episode is approximately 33 minutes long. OSNI is composed of 6 episodes, with a total of 26 hours and 22 minutes of gameplay—each episode is approximately 4 hours and 20 minutes long.

\begin{figure}[h]
 \centerline{
 \includegraphics[width=1.0\columnwidth]{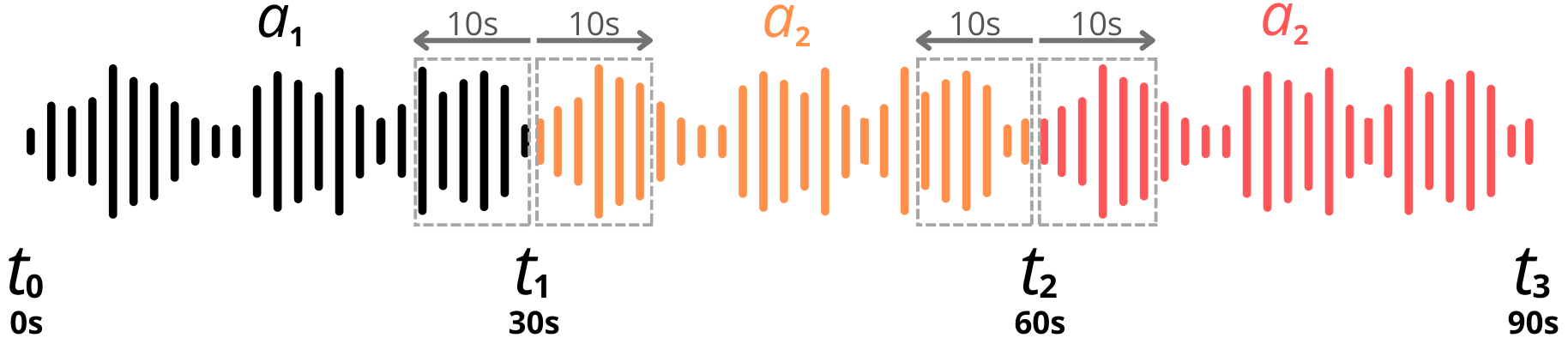}}
 \caption{The transition KLD is computed between the 10 seconds before and after every transition moment $t_i$. }
 \label{fig:transitions}
\end{figure}

We measured the performance of \bardo~with respect to three objective metrics: audio quality, alignment with the story, and transition smoothness between transcriptions. Audio quality was measured using Fréchet Audio Distance (FAD) \cite{fad}, which compares statistics computed on a set of reconstructed music clips to background statistics computed on a large set of studio-recorded music. Following the approach of Hermann1\cite{hermmann1}, we collected 32 hours of high-quality cinematic soundtracks as reference studio-recorded music. Alignment with the story was calculated with the Kullback-Leibler Divergence (KLD) with respect to the original background music of the campaigns. The transition smoothness was also computed with KLD, but by comparing the 10 seconds before and after a transition  $t_i$, as shown in Figure \ref{fig:transitions}.

In our experimental setup, we employed the YouTube API as our Speech Recognition system, extracting the transcriptions from the TRPG campaigns. We used Ollama 3.1, with 70B parameters, as \bardo's~LLM for generating the music descriptions. As our text-to-music model, we used MusicGen \cite{musicgen} large, with 3.3B parameters. For every transcription $s_i$, we generate a 30-second long audio signal $a_i$, which is the maximum length MusicGen supports. We used the VGGish model for computing FAD scores. 
We computed the KL-Divergence with the PaSST \cite{passt} classifier, which was pre-trained on the AudioSet dataset \cite{audioset}. We fine-tuned PaSST on the MTG-Jamendo dataset \cite{mtg_jamendo} to have a model more semantically suited for soundtrack classification (e.g., mood, genre, instrumentation). We fine-tuned PaSST for a single epoch with a learning rate of $10^{-4}$.

\begin{table}[t]
\centering
\setlength{\tabcolsep}{3pt}
\begin{tabular}{ccccc}

\toprule
\multicolumn{5}{c}{\textbf{$\downarrow$  Mean KL-Divergence}} \\
\toprule

\multirow{2}{*}{\textbf{TRPG}} & \multicolumn{4}{c}{\textbf{\bardo}} \\
\cmidrule{2-5}

& \multicolumn{1}{c}{\textbf{B}}   & \multicolumn{1}{c}{\textbf{E}}   & \multicolumn{1}{c}{\textbf{D}}   & 
\multicolumn{1}{c}{\textbf{DC}}  \\

\midrule

\multicolumn{1}{c}{\textbf{COTW}} & 4.84$\pm$2.98 & \textbf{3.34$\pm$1.89} & 4.26$\pm$2.65 & 4.23$\pm$2.51 \\ 

\midrule

\multicolumn{1}{c}{\textbf{OSNI}} & 5.65$\pm$3.23 & \textbf{4.16$\pm$2.12} & 4.85$\pm$2.62 & 4.96$\pm$2.86 \\ 

\bottomrule

\end{tabular}
\caption{Mean/Standard Deviation of KLD  for each \bardo~version in both COTW and OSNI.}
\label{tab:mean_kld_cotw}
\end{table}

Table \ref{tab:fad_cotw} presents the FAD audio quality metric for each \bardo~version and the original music from COTW and OSNI (Human). For each method, the FAD score is computed by retrieving a 30-minute window starting at the same random moment in both the reference background music and the generated one. This window is then split into 30-second segments. The FAD score is calculated between all generated samples of a method against the set of reference high-quality soundtracks.
Since the lower the FAD value, the 
better, \bardo- DC outperformed the other methods in both COTW and OSNI. These results suggest that conditioning MusicGen with more detailed music descriptions results in higher audio quality, both in English and Brazilian Portuguese stories.
It is important to highlight that even though \bardo- DC had a higher audio quality than the other versions, it is still not as good as professional human musical productions.

Table \ref{tab:mean_kld_cotw} shows the mean and standard deviation of the KLD story alignment metric for each \bardo~version in both COTW and OSNI. The means were calculated similarly to the FAD scores, but with slices of 10-second segments, since PaSST is limited to this context size. Moreover, each segment in the original background music had a respective segment in the generated piece. \bardo~- Emotion outperformed all other methods in both COTW and OSNI. These results indicate that the emotion of the story is a strong signal for aligning music with TRPG stories. The lower performance of \bardo~- C and \bardo~- DC is probably because Ollama can generate descriptions that trigger high-quality audio but do not necessarily align well with the story. 


Table \ref{tab:transition_kld} shows the mean and standard deviation of the KLD transition smoothness metric for each \bardo~version in both COTW and OSNI. These means were computed as in the previous metric, but comparing 10-second segments before and after a transition point $t_i$. \bardo~- Emotion outperforms all other methods in both COTW and OSNI. These results suggest that keeping a consistent music description $d_i$ with very little change over time (i.e., $d_i \approx d_{i-1}$) helps MusicGen create smooth transitions between generated audio clips $a_i$. One reason for these results might be that when the music description $d_i$ is similar to $d_{i-1}$, MusicGen focuses more on conditioning the new audio sample $a_{i}$ to the previous audio $a_{i-1}$ than to the new description $d_i$.

\begin{table}[t]
\centering
\setlength{\tabcolsep}{3pt}
\begin{tabular}{ccccc}

\toprule
\multicolumn{5}{c}{\textbf{$\downarrow$  Mean Transition KLDs}} \\
\toprule

\multirow{2}{*}{\textbf{TRPG}} & \multicolumn{4}{c}{\textbf{\bardo}} \\
\cmidrule{2-5}

& \multicolumn{1}{c}{\textbf{B}}   & \multicolumn{1}{c}{\textbf{E}}   & \multicolumn{1}{c}{\textbf{D}}   & 
\multicolumn{1}{c}{\textbf{DC}}  \\

\midrule

\multicolumn{1}{c}{\textbf{COTW}} & 2.33$\pm$2.1 & \textbf{1.33$\pm$1.19} & 2.41$\pm$2.27 & 2.19$\pm$1.93 \\ 

\midrule

\multicolumn{1}{c}{\textbf{OSNI}} & 2.11$\pm$1.65 & \textbf{1.37$\pm$1.05} & 1.88$\pm$1.47 & 2.09$\pm$1.71 \\

\bottomrule

\end{tabular}
\caption{Mean/Standard Deviation of transition KLD for each \bardo~version in both COTW and OSNI.}
\label{tab:transition_kld}
\end{table}


\section{Conclusion and Future Work}

This paper presented \bardo, a system that combines an LLM and a text-to-music model to generate background music for tabletop role-playing games. Our goal with \bardo~was to evaluate the performance of text-to-music models in long-generation tasks. We've presented four different versions of the system and evaluated them in two TRPG campaigns, one in English and another in Brazilian Portuguese. Results showed that while detailed music descriptions help improve audio quality, it is important to maintain consistency across consecutive descriptions to have smoother transitions. Moreover, emotion is a strong signal for generating soundtracks for TRPGs. 

As future work, we plan to investigate how to maintain the consistency of the generated music over time while still using detailed music descriptions. Moreover, we will conduct user studies to evaluate the quality of the generated music with subjective metrics.

\bibliography{references}

\end{document}